\documentclass[sigconf]{acmart}

\usepackage{amsthm}
\newtheorem*{problem}{Problem Description}

\usepackage{graphicx}
\usepackage{subcaption}

\usepackage{listings}
\usepackage{color}
\definecolor{mygreen}{rgb}{0,0.6,0}
\definecolor{mygray}{rgb}{0.5,0.5,0.5}
\definecolor{mymauve}{rgb}{0.58,0,0.82}
\usepackage[ruled,vlined]{algorithm2e}

\AtBeginDocument{%
  \providecommand\BibTeX{{%
    \normalfont B\kern-0.5em{\scshape i\kern-0.25em b}\kern-0.8em\TeX}}}

\setcopyright{acmcopyright}
\copyrightyear{2018}
\acmYear{2018}
\acmDOI{10.1145/1122445.1122456}

\acmConference[Woodstock '18]{Woodstock '18: ACM Symposium on Neural
  Gaze Detection}{June 03--05, 2018}{Woodstock, NY}
\acmBooktitle{Woodstock '18: ACM Symposium on Neural Gaze Detection,
  June 03--05, 2018, Woodstock, NY}
\acmPrice{15.00}
\acmISBN{978-1-4503-9999-9/18/06}

\begin{document}

\title{Connecting The Dots To Combat Collective Fraud}

\author{Mingxi Wu}
\email{mingxi.wu@tigergraph.com}
\affiliation{%
  \institution{TigerGraph Inc.}
}

\author{Xi Chen}
\email{beader.chen@gmail.com}
\affiliation{
 \institution{Information and Data Security Solutions Co., Ltd}
 }

\begin{abstract}
Modern fraudsters write malicious programs to coordinate a group of accounts to commit collective fraud for illegal profits in online platforms. These programs have access to a set of finite resources — a set of IPs, devices, and accounts etc. and sometime manipulate fake accounts to collaboratively attack the target system.
Inspired by these observations, we share our experience in building two real-time risk control systems to detect collective fraud. We show that with TigerGraph, a powerful graph database, and its innovative query language — GSQL, data scientists and fraud experts can conveniently implement and deploy an end-to-end risk control system as a graph database application. 
\end{abstract}


\keywords{graph mining, fraud detection, anomaly, graph database}


\maketitle

\section{Introduction}

Detecting fraudulent activity is a never ending battle in the digital world. More and more merchants and financial organizations are targets for fraudsters and cybercriminals. Merchants and financial services organizations will spend \$9.3 billion annually on fraud detection and prevention by 2022 \cite{juniper}. Global online payment fraud (also called CNP or “Card Not Present” fraud) alone will cost merchants \$130 billion in just five years (from 2018 to 2023) \cite{juniper2}. The latest report from LexisNexis \cite{lexis} also indicates that fraud attempts have increased significantly among retailers and e-commerce merchants during the past year, with more than twice the number of attempts and an 85 percent increase in fraud success rates. 

A large class of fraud detection systems focus on detecting fraudulent accounts, as they are the main entities which if detected and blocked in time, can save a large amount of economic loss. 

The traditional way to combat account fraud is auditing \cite{survey1}, which maintains a black list of the reported bad accounts (or tag them based on some rules) and block them when they appear. However, online fraudsters are intelligent. They usually attack an online service by using a group of accounts to achieve scale and to bypass simple fraud detection rules. These accounts sneak in a system at different time and are usually camouflaged by sequences of seemingly innocuous behavior (random logins, browsing pages etc.). When the targeted campaign is launched, they work collaboratively to attack the system. This is the so-called collective fraud \cite{hgsuspector}. 

Research on collective fraud has been focusing on devising different graph algorithms to mine account or transaction data to flag fraudulent entities \cite{hgsuspector, hitfraud, fraudar}. Graph model is widely adopted for collective fraud detection due to the collaborative nature of the attack pattern. However, most of the systems in the literature either work on an existing heterogeneous network data set \cite{hgsuspector, hitfraud} or assume there is a dedicated existing account graph data model. Little is covered to address 
the practical hurdles emerged when building an end-to-end solution given the raw online activity logs. Such issues include but not limited to how to design and represent a suitable graph model, how to maintain and handle schema evolvement, how to setup and evolve data ingestion (a.k.a data ETL) pipeline, how to implement, fine tune and maintain the data analytic queries,  how to activate the analytic query result to enable real-time risk control to prevent economic loss, and how to visually explain and further reasoning the detected fraudulent patterns. We point out that all these practical problems if handled in isolation without a standard data representation and programming language at different stages, it demands diverse background data engineers to collaborate across stages, adding non-trivial communication overhead and venues for errors. 

In this paper, we share our experience in building a real-time risk control system to prevent collective fraud end-to-end. We adopt TigerGraph \cite{tgmpp}, a native MPP graph database as our back-end server. The benefits of using a native graph database are unified data representation and programming API, data independency of application development and transparent query optimization. 

Section 2 and 3 presents the background information of TigerGraph architecture and GSQL. Section 4 describe our framework of the risk control system. Section 5 and 6 detail two applications. The paper is concluded in Section 6. 

\section{Background}
\begin{figure}[ht]
  \centering
  \includegraphics[width=1 \linewidth]{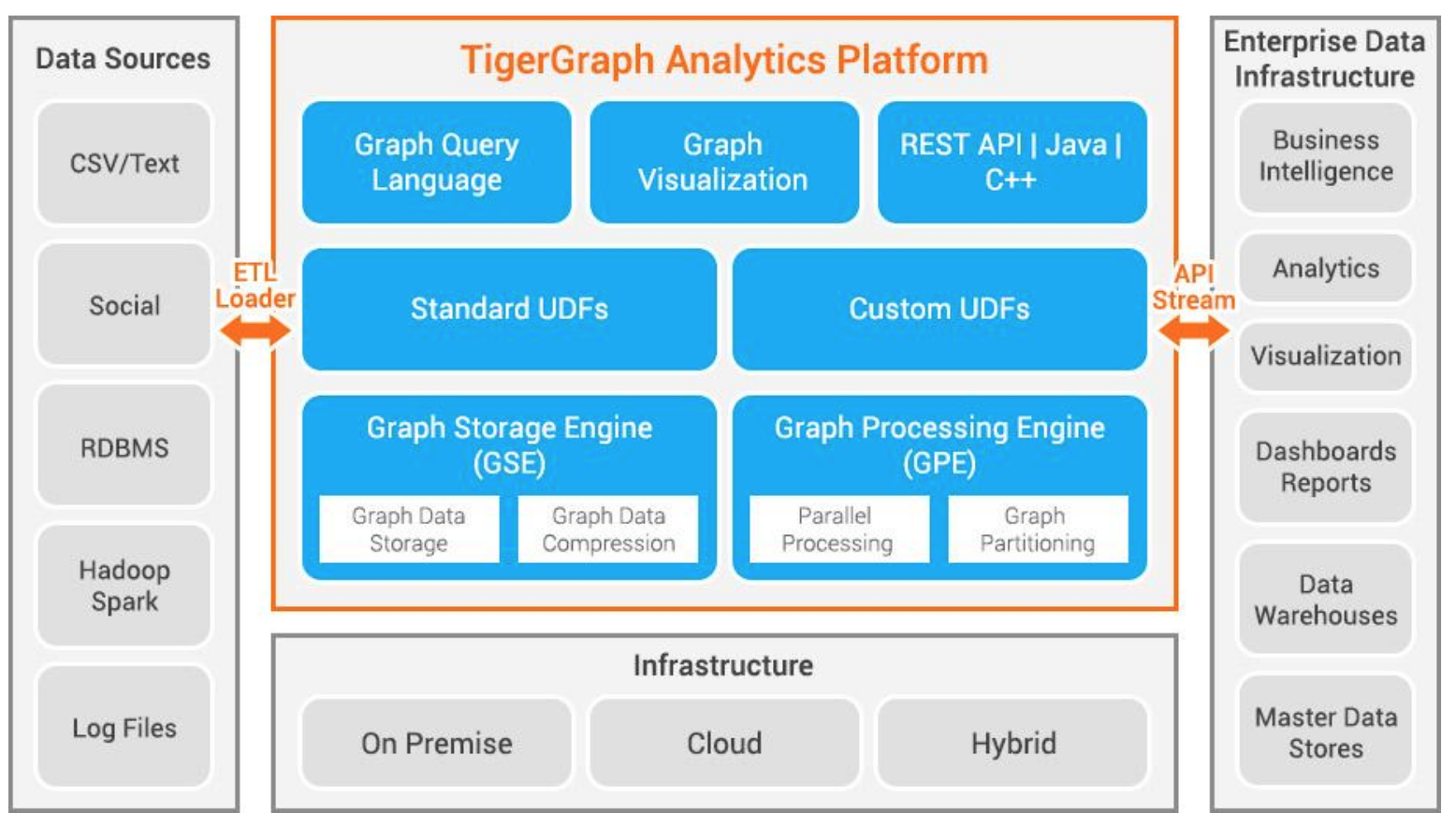}
  \caption{TigerGraph System Architecture}
  \label{fig:architecture}
\end{figure}

TigerGraph’s architecture is depicted by Figure \ref{fig:architecture} (in the blue boxes). The system vertical stack is structured in three layers: the top layer comprises the user programming interface,
which includes the GSQL compiler, the GraphStudio visual SDK, and REST APIs for other languages to connect to TigerGraph; the middle layer contains the standard built-in
and user defined functions (UDFs); the bottom layer includes the graph storage engine (GSE) and the graph processing engine (GPE). We elaborate on GSQL as it's the programming API to use TigerGraph. 

GSQL is a unified declarative graph database language for users to build applications end-to-end. It covers data modeling, data loading and data querying. 

\smallskip
\noindent \textbf{DDL.}
GSQL’s DDL shares SQL’s philosophy of defining in the same CREATE statement a persistent data container as well as its type. This is in contrast
to typical programming languages which define types as standalone entities, separate from their instantiations.

GSQL’s CREATE statements can define vertex containers, edge containers, and graphs consisting of these containers. The attributes for the vertices/edges populating the container are declared using syntax borrowed from SQL’s CREATE TABLE command. In TigerGraph’s model, a graph may contain
both directed and undirected edges. 

Listing \ref{ddl} shows an example of DDL, modeling an incentive campaign of a company. This campaign encourages its user to invite their friends to register an account. Bonus will be given to each account inviting new accounts.   

\begin{lstlisting}[language=SQL, caption={DDL Example}, label={ddl}]
CREATE VERTEX Account (id STRING PRIMARY KEY, phone STRING, reg_time DATETIME)
CREATE VERTEX IMEI (id STRING PRIMARY KEY, imei STRING) 
CREATE VERTEX Order (id STRING PRIMARY KEY, order_date DATETIME)
CREATE UNDIRECTED EDGE use_imei (FROM Account, TO IMEI)
CREATE DIRECTED EDGE invite (FROM Account, TO Account)
CREATE DIRECTED EDGE send_bonus (FROM Account, TO Order)
CREATE DIRECTED EDGE recv_bonus (FROM Order, TO Account)
CREATE GRAPH IncentiveCampaignGraph (*)
\end{lstlisting}

\begin{lstlisting}[language=SQL, caption={DDL Example}, label={dll}]
USE GRAPH IncentiveCampaignGraph
CREATE LOADING JOB load_orders {
    DEFINE FILENAME jsonfile;
    LOAD jsonfile
      TO VERTEX BonusOrder VALUES($"order_id", "order_date"),
      TO EDGE send_bonus VALUES($"sendr_phone", $"order_id"),
      TO EDGE recv_bonus VALUES($"order_id", $"recvr_phone"),
    USING JSON_FILE="true";
}
CREATE LOADING JOB load_invite {
    DEFINE FILENAME invite_file;
    LOAD invite_file
      TO EDGE invite VALUES($0, $1);
}
RUN LOADING JOB load_orders USING jsonfile="/home/orders.json"
RUN LOADING JOB load_invite USING invite_file="/home/invite.csv"
\end{lstlisting}
\smallskip
\noindent \textbf{DLL.}
GSQL provides a declarative loading language (DLL) to allow user to specify a loading job. The loading job can be invoked offline and online. 
The DLL is a mapping language. It comprises one or multiple LOAD statements. Each LOAD statement maps a source file schema to the targeted vertex or edge schema. 
The source file can be a Kafka source, or it can be raw file in JSON or CSV format. User references the source file column by its name or its column position in the VALUES clause. This is in resemblance of SQL INSERT statement. If column reference by position is used, the position numbering always starts from 0 and is represented by \$0.

For each loading job defined, a REST endpoint is created to allow online ingestion. For offline batch file loading, file binding happens at job invocation time, where "RUN LOADING JOB" command provides a USING clause to allow flexible run-time file binding. Listing \ref{dll} shows an example of DLL. 

\smallskip
\noindent \textbf{DML.}
GSQL DML is the core of the query language. The guiding principle behind the design of GSQL was to facilitate adoption by SQL programmers while simultaneously flattening the learning curve for novices, especially for adopters of the BSP programming paradigm \cite{bsp}.

To this end, GSQL’s design starts from SQL, extending its syntax and semantics parsimoniously, i.e. avoiding the introduction of new keywords for concepts that already have
an SQL counterpart. We summarize the key additional primitives. Please check \cite{gsql101} for more detailed description.

Each GSQL query can be installed as a stored procedure and invoked via a REST endpoint. 

\begin{itemize}
    \item \textbf{Graph Patterns} in the FROM Clause. GSQL extends SQL’s FROM clause to allow the specification of patterns. Patterns
specify constraints on paths in the graph, and they also contain variables, bound to vertices or edges of the matching paths. In
the remaining query clauses, these variables are treated just like standard SQL tuple variables.
     \item \textbf{Accumulators} are special data types for GSQL developer to use.  Local accumulator (prefixed by "@") serves as vertex's run-time attribute. Global accumulator (prefixed by "@@" stores the run-time query state. The aggregation results can be distributed across vertices, to support multi-pass and, in conjunction with loops, even iterative
graph algorithms implemented in MPP fashion.
   \item \textbf{Flow controls} are supported in GSQL,  in particular loops, which are essential to support standard iterative graph
analytics (e.g. PageRank \cite{pagerank}, shortest-paths \cite{algbook}, weakly connected components \cite{algbook}, recommender systems, etc.).

\end{itemize}

\begin{figure}[ht]
  \centering
  \includegraphics[width=1 \linewidth]{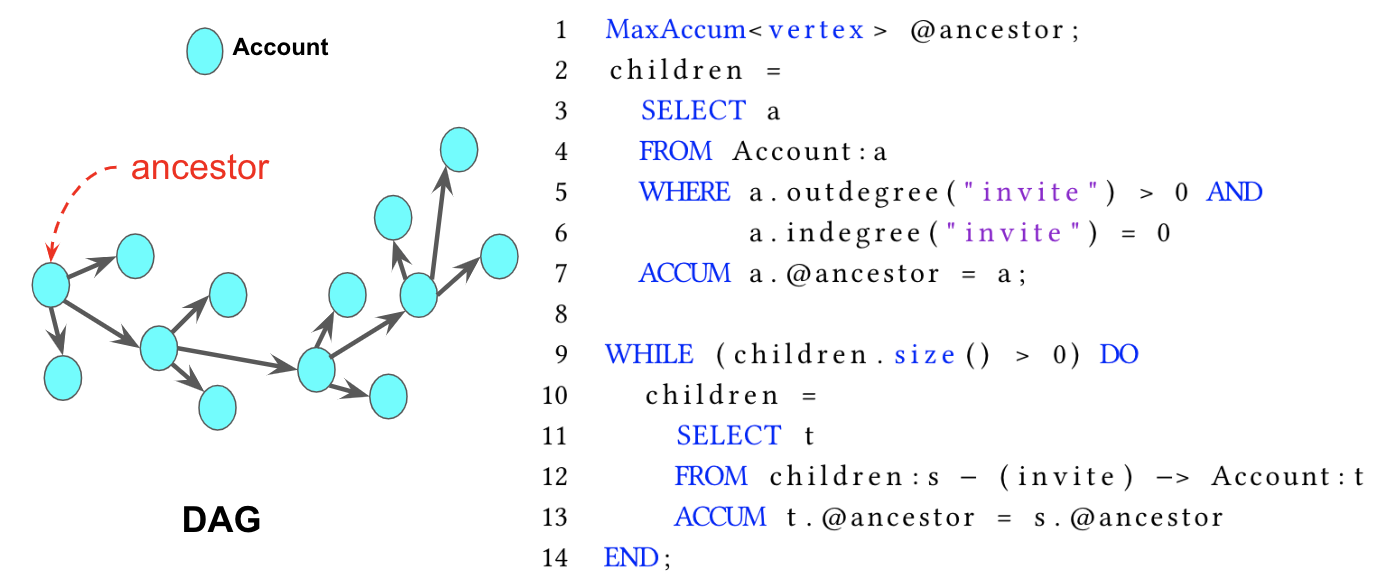}
  \caption{Find Connected Component Using GSQL}
  \label{fig:intro}
\end{figure}

We use an example to show GSQL. The left of Figure \ref{fig:intro}  illustrates a pattern that we discovered from an online incentive campaign log. To achieve the networking effect of marketing, the merchant launched an incentive campaign to reward their existing users for inviting others to register. We notice this invitation chain is very long, more than 60 hops in depth, and each invitor in this chain invited exactly the same number of invitees, reflecting mechanic behavior.

The invitation graph is a directed acyclic graph (DAG). We discovered this pattern by running a connected component (CC) algorithm coded using the GSQL query on the right of Figure \ref{fig:intro}. Line 1 declares a vertex runtime attribute (accumulator) named @ancestor. Its type is MaxAccum<vertex>, a builtin accumulator type,  which has an initial value, and the user can keep accumulating (using its “+=” built-in combiner operator) new values into it. The name MaxAccum suggests this accumulator maintains the maximum vertex id it ever sees. Line 2-7 is the first query block,  in the form of SELECT-FROM-WHERE-ACCUM structure. It matches all Account vertices (FROM clause) and finds those (WHERE clause) that has zero incoming edge and non-zero outgoing edges; these are  CC ancestors. The ACCUM clause updates the ancestors’ run-time attribute @ancestor with its own id. 

After we find the CC ancestors, the second query block (Line 10-13) propagates the ancestors from the current vertex set to their direct children accounts. The edge pattern is specified in the FROM clause, where s and t are two bind variables at the source and the target of an edge, respectively. The ACCUM clause relays the source ancestor id to the target vertex. The SELECT clause assigns all the target vertices to a vertex set variable “children” (Line 10). The WHILE loop (Line 9) iterates the second query block until no more descendants are discovered.  In the end, each vertex of a CC with size greater than 1 is adorned with their CC ancestor id in its @anecstor attribute. 

Once we find the CCs, we can profile them and report those that have long chains as anomalies. Or,  we can calculate the CC depth while propagating its ancestors.

This example showcases the conciseness and convenience of using GSQL \cite{tgmpp}.

\section{Risk Control System}
In this section, we give a high level overview of our risk control system architecture. 

\begin{problem}
Given the account activity logs, the goal is to continuously detect collective fraud accounts, mark them with risk scores, and allow user to query their risk scores.  
\end{problem}

\begin{figure}[ht]
  \centering
  \includegraphics[width=1 \linewidth]{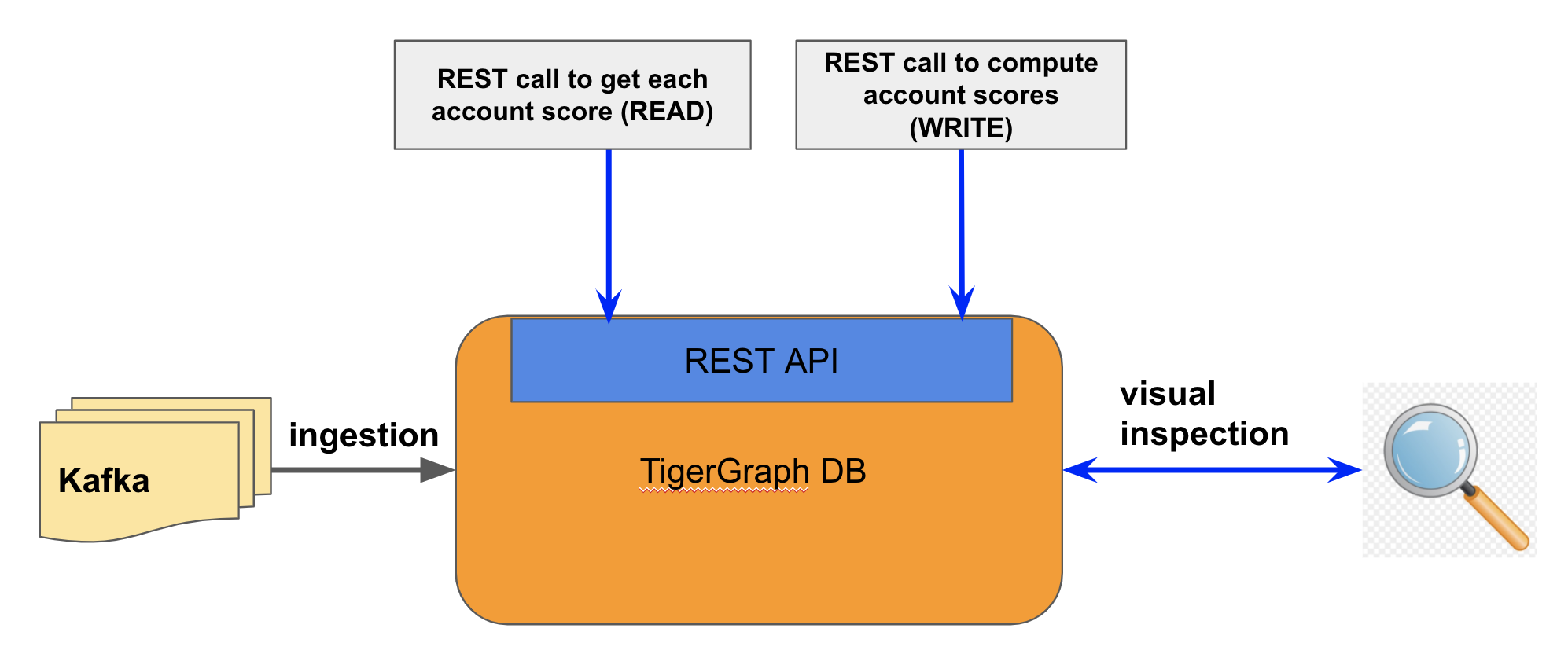}
  \caption{Risk Control System Architecture}
  \label{fig:risk}
\end{figure}
An account is a logic concept. It represents a digital entity who is the first class citizen in an online service. Usually, each account is associated with one natural person. However, fraudsters are interested in using programs to manipulate a pool of fake or stolen accounts to simulate human behavior and obtain illegal profits. 

The whole architecture is shown in Figure \ref{fig:risk}. We use TigerGraph as the backend server to manage a dedicated account graph for fraud detection. The application designers first design an account graph model using GSQL's DDL. Next, the account activity logs are real-time bufferred in a Kafka cluster, and ingested into the account graph using a GSQL's DLL defined loading script. GSQL DML queries are used to code any fraud detection algorithms and point look-up query. All queries can be installed as a service and invoked via a REST endpoint.  Via TigerGaphs's visual SDK, user can visually explore the subgraphs surrounding a risky account. 

Since the collaborative fraud accounts will form a connected component (CC) when they collaborate or share resources (IPs, Devices etc.), we adopt the following framework to detect collective fraud. 
\begin{enumerate}
    \item An account graph is first constructed using either explicit or implicit relationships. Example, account-invite-account is an explicit relationship; an account shares an IP with another account in a time window is an implicit relationship. 
    \item Next, a set of analytical GSQL queries (running periodically) will find account CCs and profile each CC using an aggregate statistic. 
    \item The accounts in the CCs with anomalous statistics are flagged as fraudulent accounts. 
\end{enumerate}
\begin{figure*}[ht]
\begin{minipage}{0.88\linewidth}
\fbox{
  \centering
  \begin{subfigure}[b]{0.45\linewidth}
    \includegraphics[width=\linewidth]{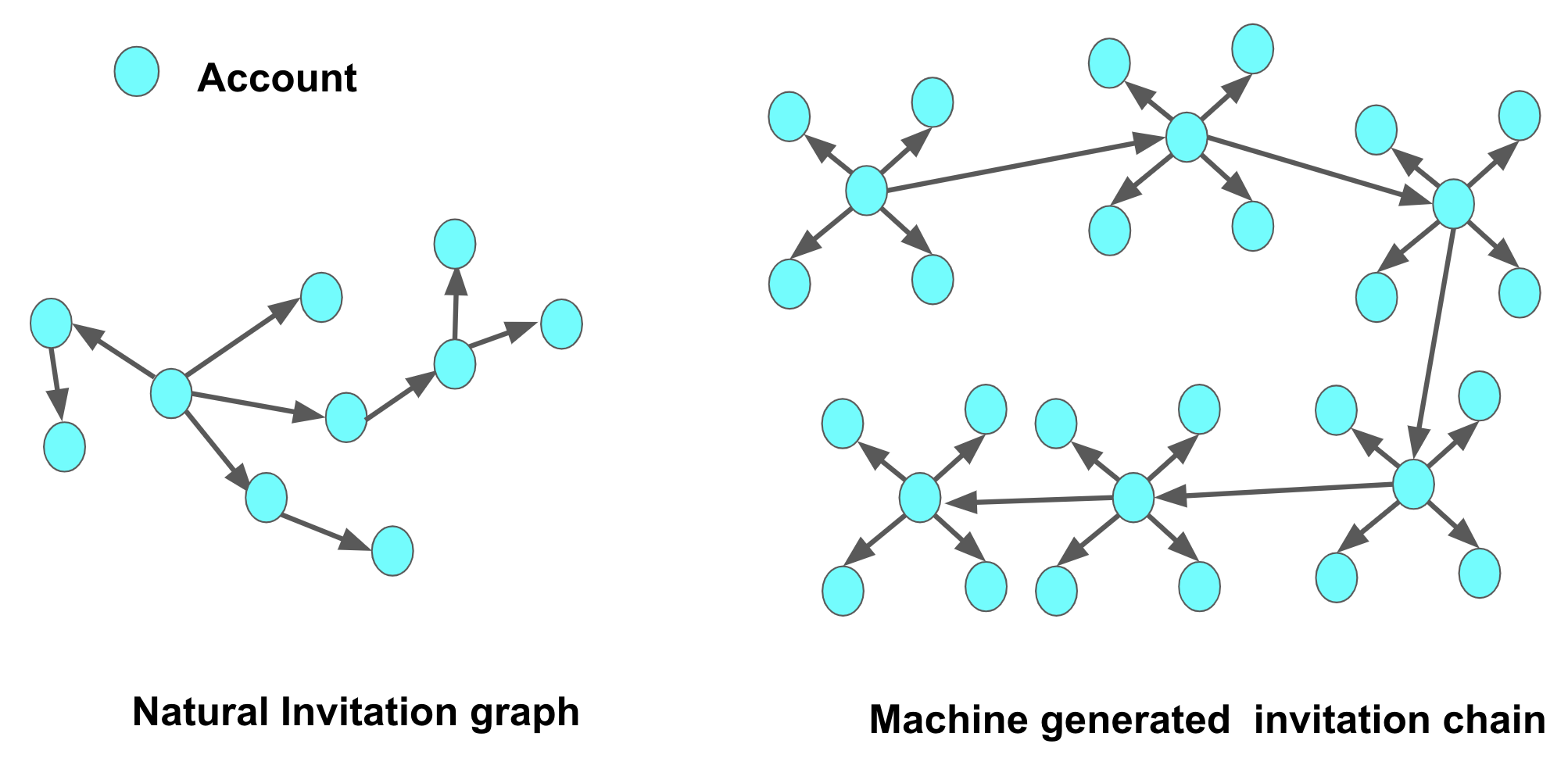}
    \caption{CC Comparison by Depth}
  \end{subfigure}
  }
  \fbox{
\begin{subfigure}[b]{0.46\linewidth}
    \includegraphics[width=\linewidth]{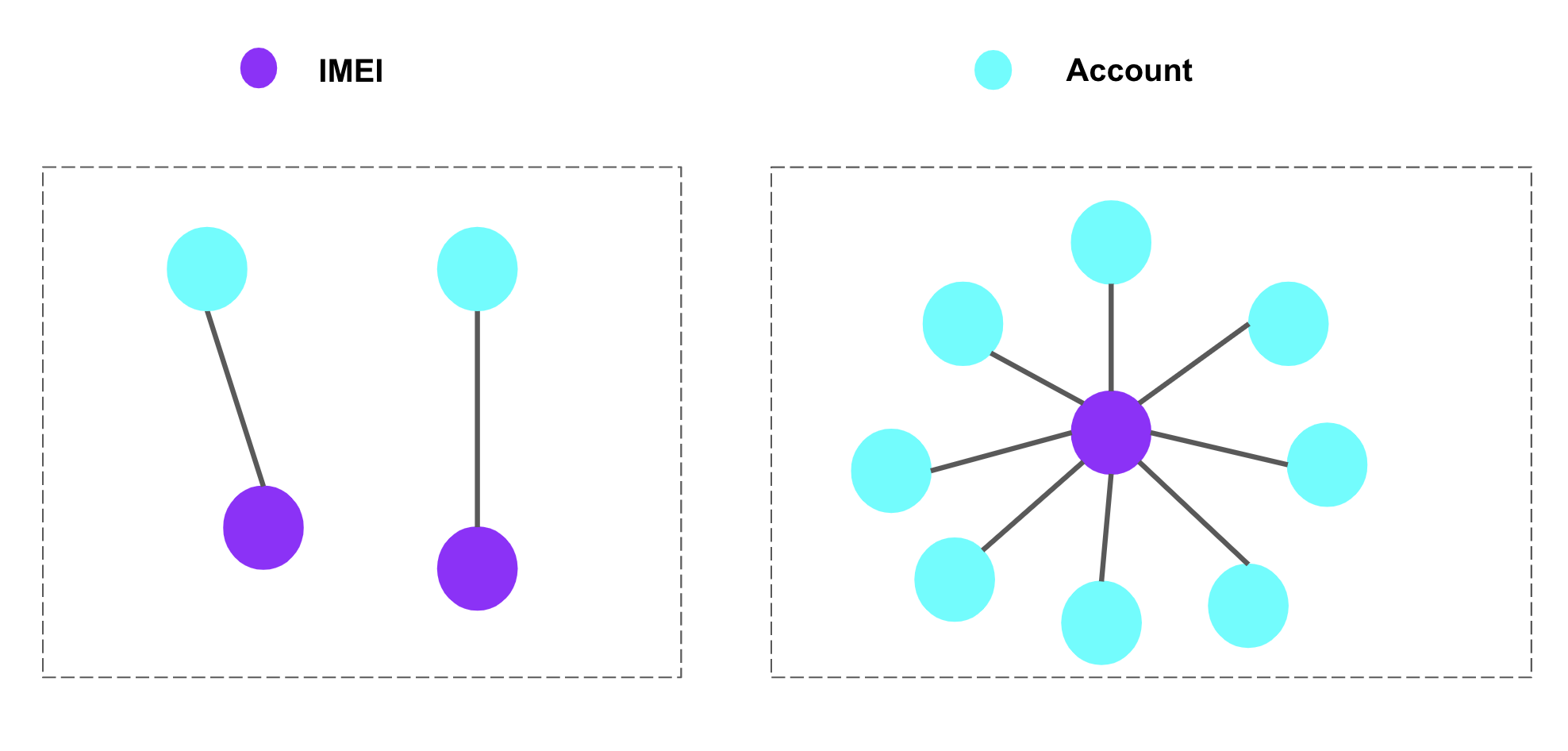}
    \caption{CC Comparison by Shared-device-ratio}
  \end{subfigure}
  }
\end{minipage}
\begin{minipage}{0.88\linewidth}
\fbox{
  \centering
  \begin{subfigure}[b]{0.45\linewidth}
    \includegraphics[width=\linewidth]{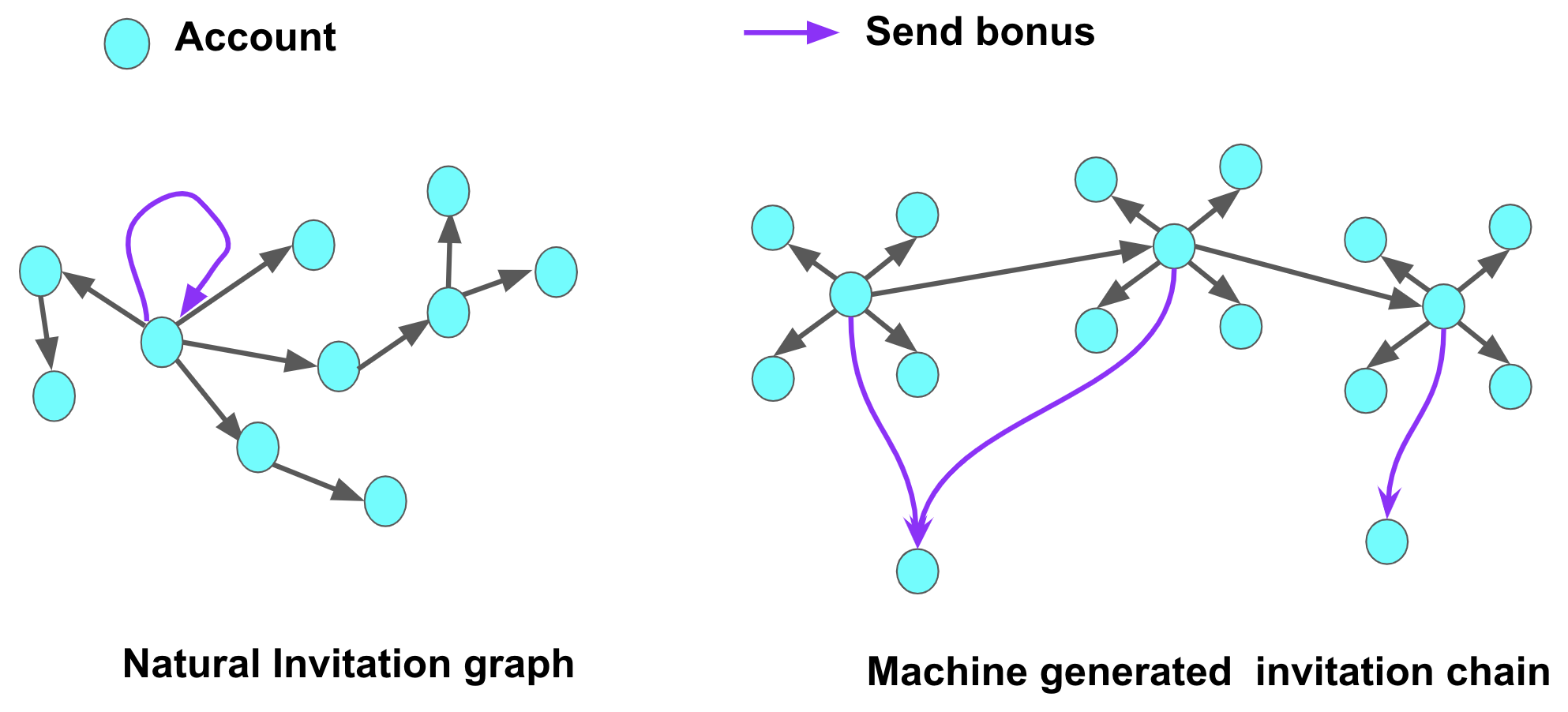}
    \caption{CC Comparison by Non-self-order-ratio}
  \end{subfigure}
  }
  \fbox{
\begin{subfigure}[b]{0.46\linewidth}
    \includegraphics[width=\linewidth]{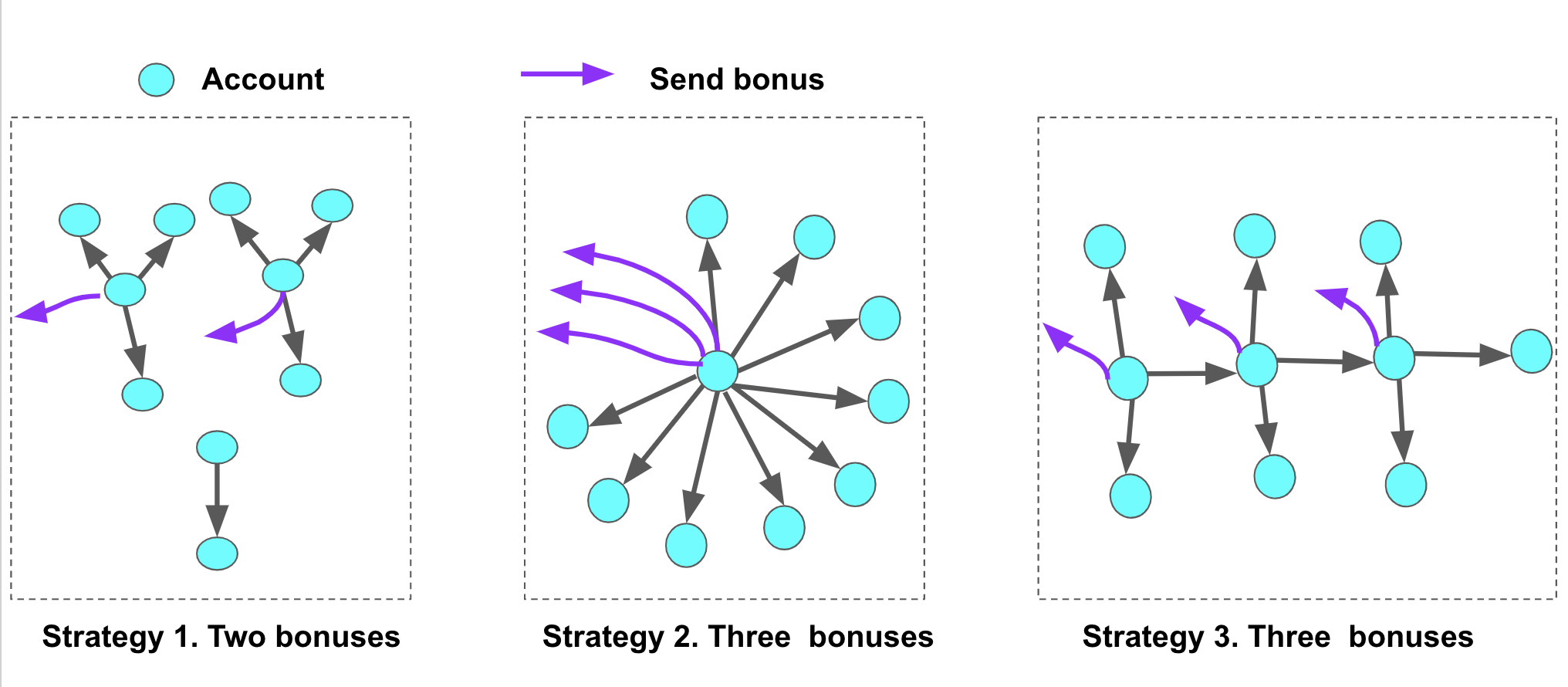}
    \caption{CC Comparison by Gini-index}
  \end{subfigure}
  }
  
  \caption{Profiling CCs}
\label{fig:CC}
\end{minipage}
\end{figure*}

\section{Application I}

\begin{table}
  \caption{Invitation Graph Statistics.}
  \label{Tab:data1}
  \begin{tabular}{llll}
    \toprule
    \textbf{Vertex} & \textbf{Count} & \textbf{Edge} & \textbf{Count}\\
    \midrule
    Account & 93237 & use\_imei & 24623 \\
    \midrule
    Order & 19096 & send\_bonus & 19090 \\
    \midrule
    IMEI & 19833 & recv\_bonus & 18724 \\
    \midrule
      &  & invite & 74354 \\
        \bottomrule
  \end{tabular}
\end{table}
In this application, we explore the explicit relationships between two accounts and between accounts and their shared resources to construct our account graph.

\subsection{Case Description}
A company launched an incentive campaign, which encourages its user to invite their friends to open a new account with the company. Bonus will be given to the account inviting 10 new accounts. Soon after the campaign was launched, the company noticed that some accounts who won bonus sent their credits to one beneficiary account.  This behavior indicates unusual user pattern.  The goal here is to detect those accounts that are manipulated by fraudsters to illegally win the referral bonus. 

There are three logs: the bonus order log, the phone device log, and the referral log. The bonus order log schema is (order\_id, order\_date, bonus\_name, sender\_phone, recv\_phone). The phone device log schema is (phone\_number, imei). The referral log has the schema (recv\_phone, recv\_reg\_date, sender\_phone, sender\_reg\_date). 

\subsection{Invitation Graph}
In this graph construction, we use the explicit invite relationship to connect accounts. And we connect each account with their order and IMEI to explore their shared resource pattern. 
Specifically, we create the Account, the Order, the IMEI vertex types, the un-directed use\_imei(connecting an Account and an IMEI), the directed invite(connecting two Accounts), and the directed  send\_bonus (connecting an Account and an Order) and recv\_bonus (connecting an Account and an Order) edge types.

We model and map the three logs to the graph schema  using GSQL illustrated in Listing \ref{ddl} and Listing \ref{dll} respectively. 

\subsection{Analytical Queries}
The analytic query we use is to cluster accounts into CCs using only the "invite" edges. After we find the account CCs, we profile each account CC with a statistic. Four CC profiling statistics are used in this application. 
\begin{itemize}
	\item \textbf{CC Depth.} In Fig. \ref{fig:CC} (a), the left shows a natural invitation graph, where an invitee has equal chance of inviting others or do not invite others. The right shows a machine manipulated pattern where accounts are re-used as much as possible, forming a long chain of invitees.
	\item \textbf{Average Account Number Per Device.} In Fig. \ref{fig:CC} (b), the left shows one IMEI usually identifies one account. The right shows fraudsters use one IMEI to register multiple accounts, which can be caught by the averaged accounts per IMEI of an account CC. 
	\item \textbf{Non-self-order Ratio.} In Fig. \ref{fig:CC} (c), the left shows a normal behavior where an invitor sends the claimed bonus to itself. The right shows a fraudulent behavior where invitors collectively aggregates the claimed bonuses to one or multiple beneficiaries, which can be detected by the percentage of accounts in each CC that send their bonus to others.
	\item \textbf{Gini index.} The Gini index \cite{gini} measures the inequality among values of a frequency distribution. We use the Gini index of each account invited guest number to capture collective fraud. Fig. \ref{fig:CC} (d) shows why fraudsters will likely form a homogeneous invitation chain.  Three scenarios all use 10 accounts each. Assuming 3 invitees can help their invitor win 1 bonus. To maximize the claimed bonus, (d) shows three  strategies. Strategy 1 wins 2 bonus. Strategy 2 and strategy 3 win 3 bonus each. However, strategy 2 is easily to be blocked if the system requires that one account at most claims 1 bonus. Strategy 3 not only by pass the aforementioned simple blocking rule, it also maximizes the utilization of the 10 accounts. 
\end{itemize}

\subsection{Experiment}
\smallskip
\noindent \textbf{Goal.} Our goal is to answer two questions. 

\begin{enumerate}
 \item What are the precision and recall using this method? 
 \item Each CC statistics targets a hypothetical collective fraud pattern, do these patterns exist in a real data?
\end{enumerate}

\begin{figure}[ht]
  \centering
  \includegraphics[width=1 \linewidth]{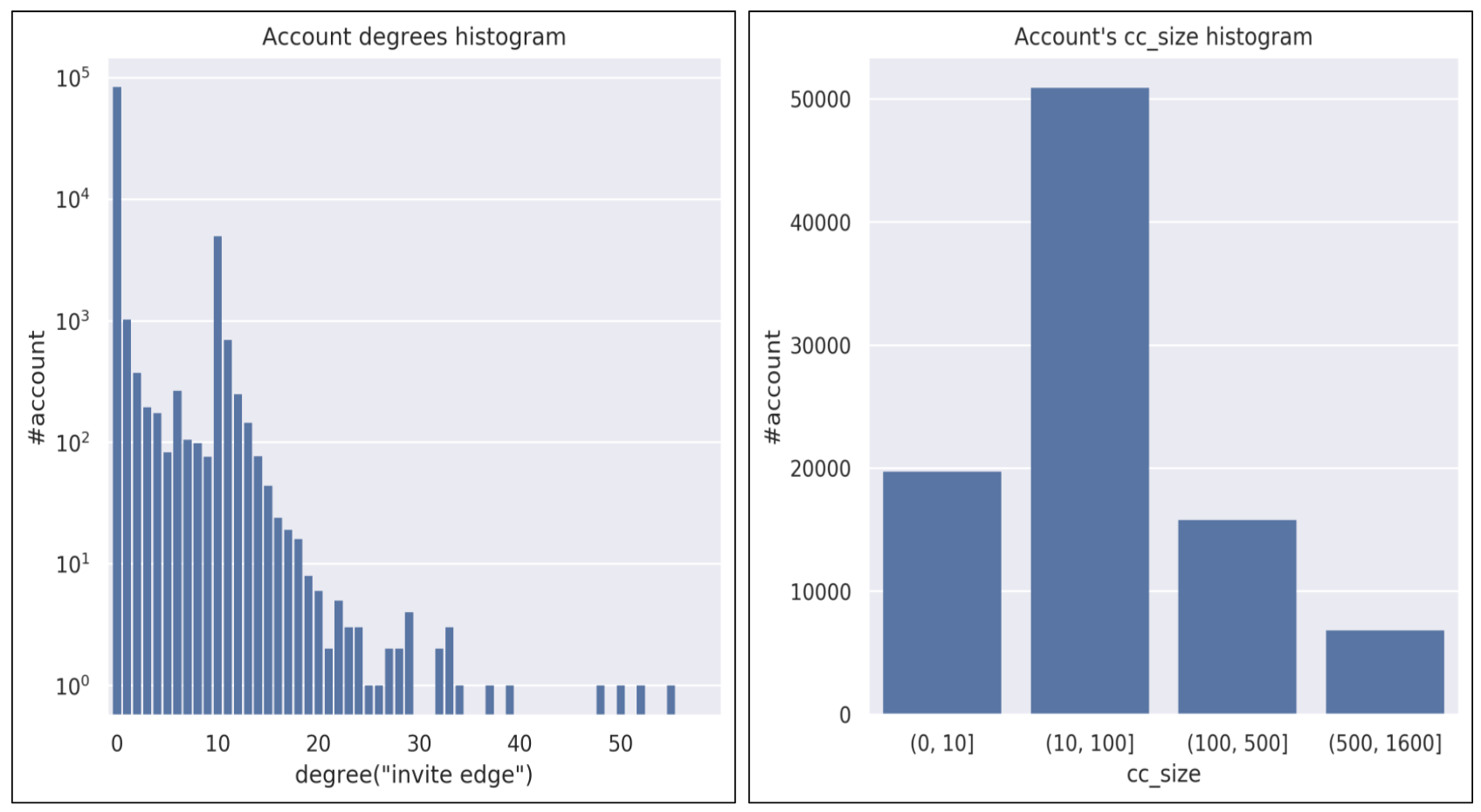}
  \caption{Invitation Graph Statistics.}
  \label{fig:data1}
\end{figure}

\smallskip
\noindent \textbf{Setup.} This is a proof of concept (PoC) project conducted for our client using the post-campaign data. We applied the system over 1 month real data after the campaign was launched. We computed the CC using the "invite" edge only, and finally used the 4 CC statistics to detect anomalies.  

Table \ref{Tab:data1} shows the vertex and edge statistics for each type. 

Figure \ref{fig:data1} shows the invitation graph's statistics. In general, the account out-degree histogram (the left chart) follows the power-law with the exception that there is a spike in the bucket positioned at the out-degree value 10.  The campaign manager shared with us that this campaign has set the invitees threshold number to be 10 for an invitor to win a referral bonus, which explains the spike value. The account distribution in different CC size histogram (the right chart) shows that more than half of the accounts are in a CC whose size is between 10 and 100. 

\begin{table}[]
\begin{tabular}{|l|l|l|l|l|}
\hline
Alg & Description                                                                                                               & \#Account & \begin{tabular}[c]{@{}l@{}}\#Bonus\\ Order\end{tabular} & \#CC \\ \hline
a   & cc\_depth \textgreater 5                                                                                                  & 18735     & 1691                                                    & 103  \\ \hline
b   & \begin{tabular}[c]{@{}l@{}}cc\_bonus\_sent\textgreater{}10,\\  cc\_non-self-receiver\_ratio\textgreater{}0.5\end{tabular} & 12653     & 1090                                                    & 59   \\ \hline
c   & \begin{tabular}[c]{@{}l@{}}cc\_size\textgreater{}=30, \\ cc\_shared\_device\_ratio\textgreater{}2\end{tabular}            & 3377      & 334                                                     & 11   \\ \hline
d   & \begin{tabular}[c]{@{}l@{}}cc\_size\textgreater{}=30, \\ gini\_index\textless{}0.1\end{tabular}                           & 20920     & 1848                                                    & 257  \\ \hline
e   & \#account\_per\_imei\textgreater{}=3                                                                                      & 4763      & 2869                                                    & NA   \\ \hline
f   & \#senders\_per\_receiver\textgreater{}=3                                                                                  & 2391      & 2464                                                    & NA   \\ \hline
g   & a+b+c+d                                                                                                                   & 30054     & 2648                                                    & 289  \\ \hline
h   & e+f                                                                                                                       & 5718      & 4396                                                    & NA   \\ \hline
i   & g+h                                                                                                                       & 33346     & 5430                                                    & NA   \\ \hline
j   & g+h-d                                                                                                                     & 25760     & 5071                                                    & NA   \\ \hline
\end{tabular}
\caption{Different CC Methods}
\label{Tab:cc}
\end{table}

We wrote a GSQL DDL script define the schema of the invite graph. Another GSQL DLL script was defined to load the three logs into the invite graph. The four CC profiling algorithms were written in GSQL DML, and extending the basic DAG CC detection algorithm depicted in Figure \ref{fig:intro}. For each CC profiling algorithm, we pick a very conservative threshold parameter to detect anomaly. These threshold are specified in the Description column of Table \ref{Tab:cc}.  All the GSQL scripts can be accessed online\footnote{Incentive Fraud GSQL scripts: \url{http://github.com/beader/combat_collective_fraud}.}.

To evaluate the precision and recall of our method, we need to know which accounts are true fraud accounts. We rely on Tencent T-Sec product, which is an anti-fraud cloud service to provide the risk score for each phone number. We rely on T-Sec to flag our test data since Tencent has billions of monthly active users and more data to profile each user.

\smallskip
\noindent \textbf{Result.}  Table \ref{Tab:cc} presents the results. (1) This campaign gave out \$112,798 incentive bonus. Conservatively estimate, 33346 accounts are fraudsters controlled, they collectively claimed 6000 bonus. Each bonus claim costs \$7 loss to our client, total loss added up to \$42,299.40. (2) Exclude Gini index method (which was calculated based on each invitor's invitee count in a CC), still we have 25,760 fraudulent accounts, which claimed 5071 incentive bonus orders, amounting to \$40,184. (3) When calculate anomalous accounts based on e and f, we do not use CC. They are one step statistics of an account. We notice the discovered accounts is significantly less than using the CC methods. From calling Tecent APIs, we compute that the combined methods achieved the precision of 94.7\% and the recall of 81.9\%. 

\begin{figure*}[ht]
\begin{minipage}{1\linewidth}
\fbox{
  \centering
  \begin{subfigure}[b]{0.445\linewidth}
    \includegraphics[width=\linewidth]{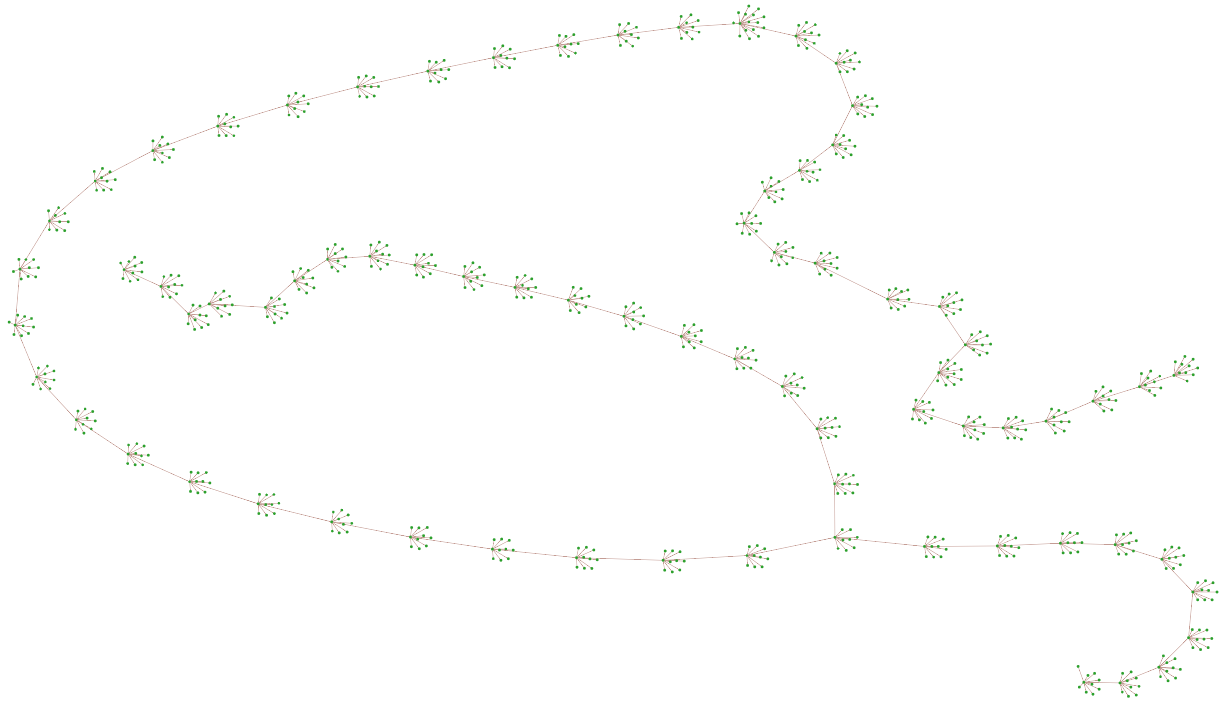}
    \caption{Deep-depth CC}
  \end{subfigure}
  }
  \fbox{
\begin{subfigure}[b]{0.445\linewidth}
    \includegraphics[width=\linewidth]{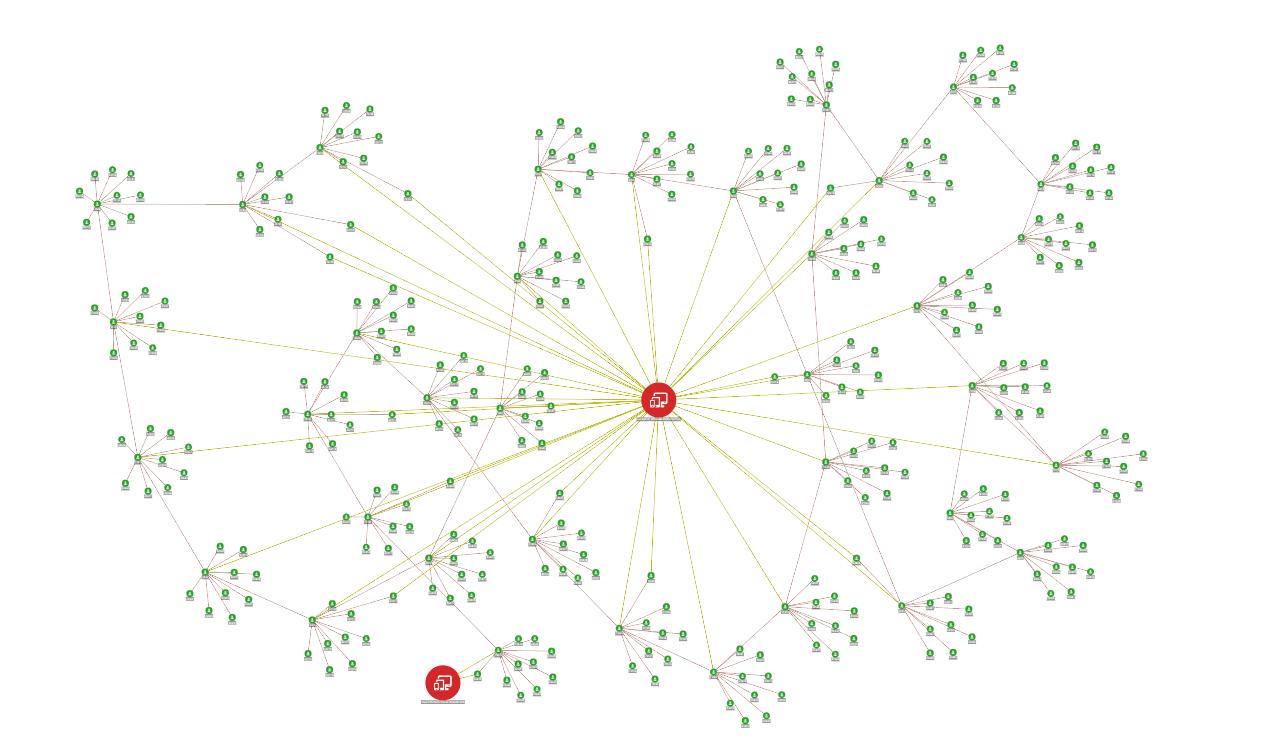}
    \caption{High Shared-device-ratio CC}
  \end{subfigure}
  }
\end{minipage}
\begin{minipage}{1\linewidth}
\fbox{
  \centering
  \begin{subfigure}[b]{0.445\linewidth}
    \includegraphics[width=\linewidth]{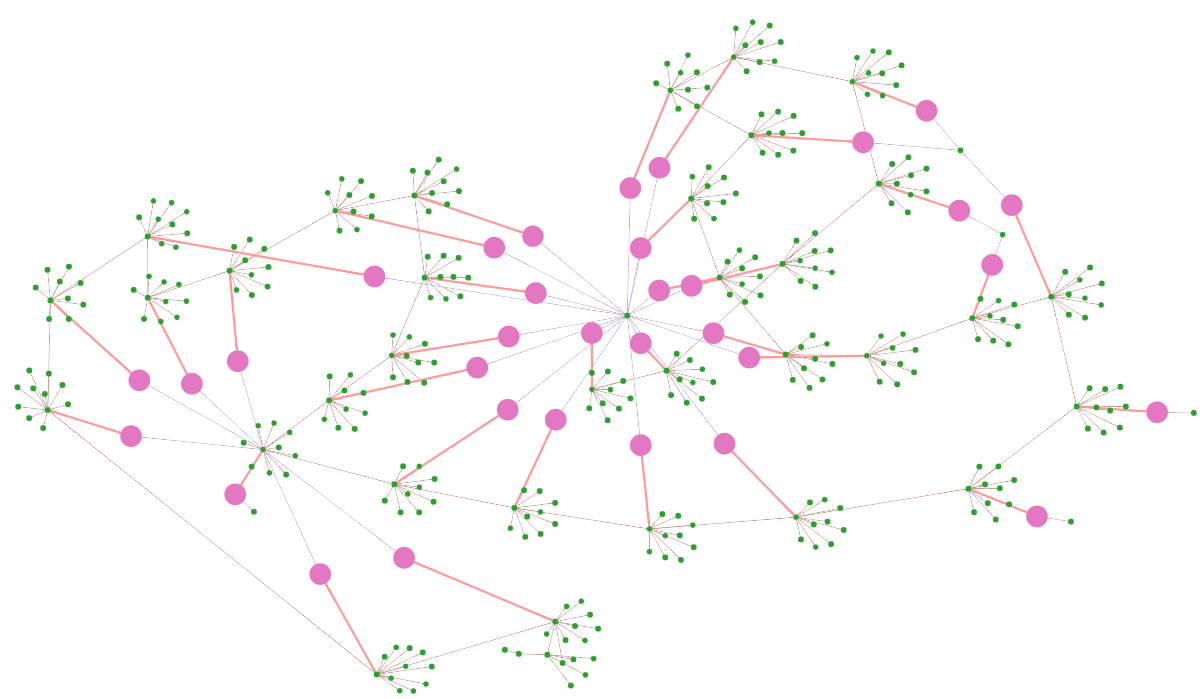}
    \caption{High Non-self-order-ratio CC}
  \end{subfigure}
  }
  \fbox{
\begin{subfigure}[b]{0.45\linewidth}
    \includegraphics[width=\linewidth]{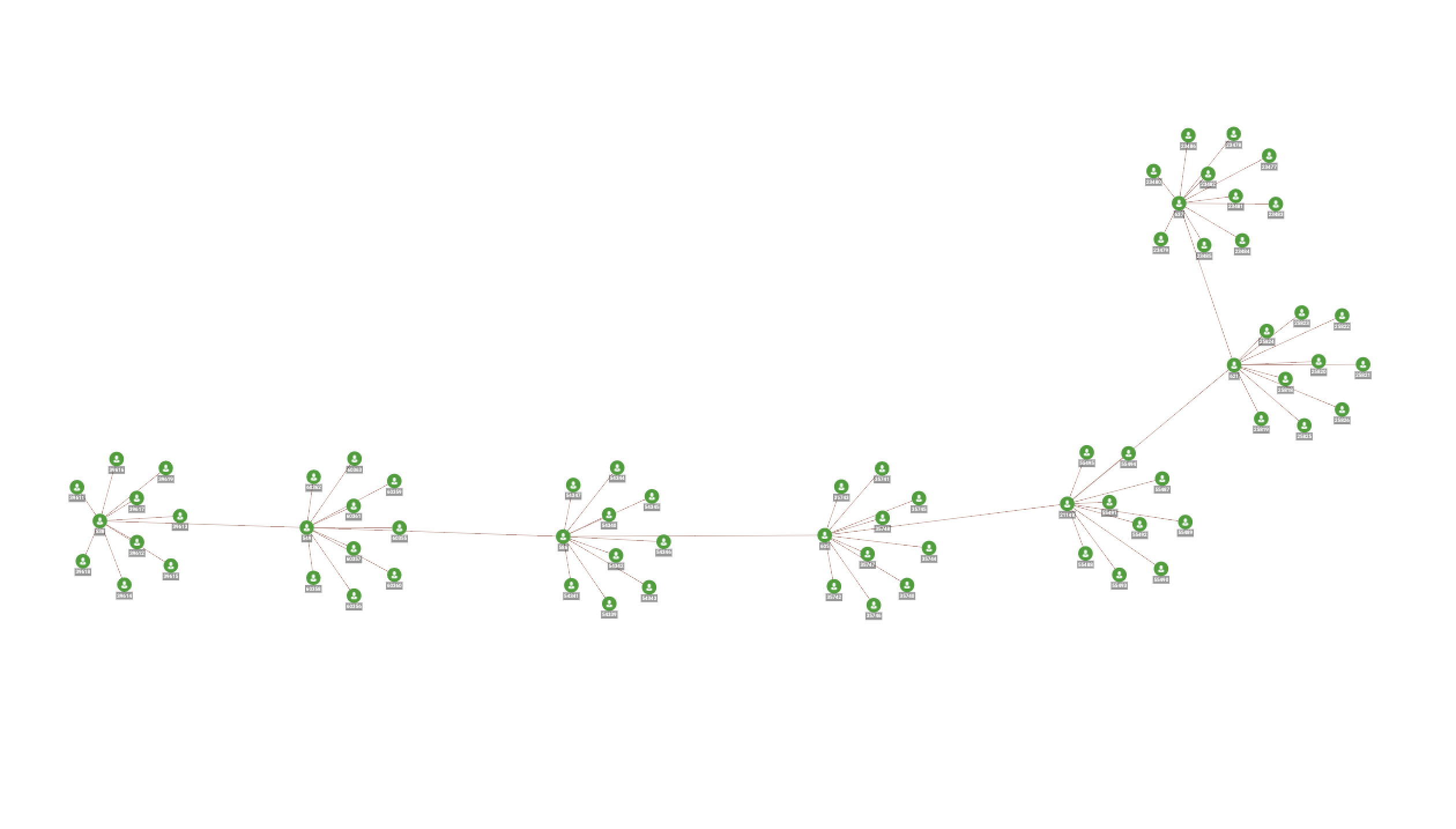}
    \caption{Low Gini-index CC}
  \end{subfigure}
  }
  
  \caption{CC Patterns of Different Metric On A Real Data Set}
\label{fig:CC-real}
\end{minipage}
\end{figure*}

Further, we show some examples of the verified fraudulent CCs  in Figure \ref{fig:CC-real}. They are explainable following our initial CC statistics design hypothesis.(a) and (b) are self-explanatory. In (c) we show a CC that all the bonus claims were sent to others. The pink vertices represent a claim transaction, which connects the sender and receiver phone. It shows some claims were sent to one account that does not belong to the CC (the central one), and some claims were aggregate to some account within the CC (the left one). In (d), we show a low Gini index CC, where each phone  invited 10 phones, 9 surround the invitor, and 1 forms a new invitee cluster. 

Based on the positive result of the PoC, the client plans to deploy the detection framework for their next campaign, running CC profiling periodically, and raise alarm when discovering CC that outnumber the pre-set cutoff.
\begin{figure}[ht]
  \centering
  \includegraphics[width=1 \linewidth]{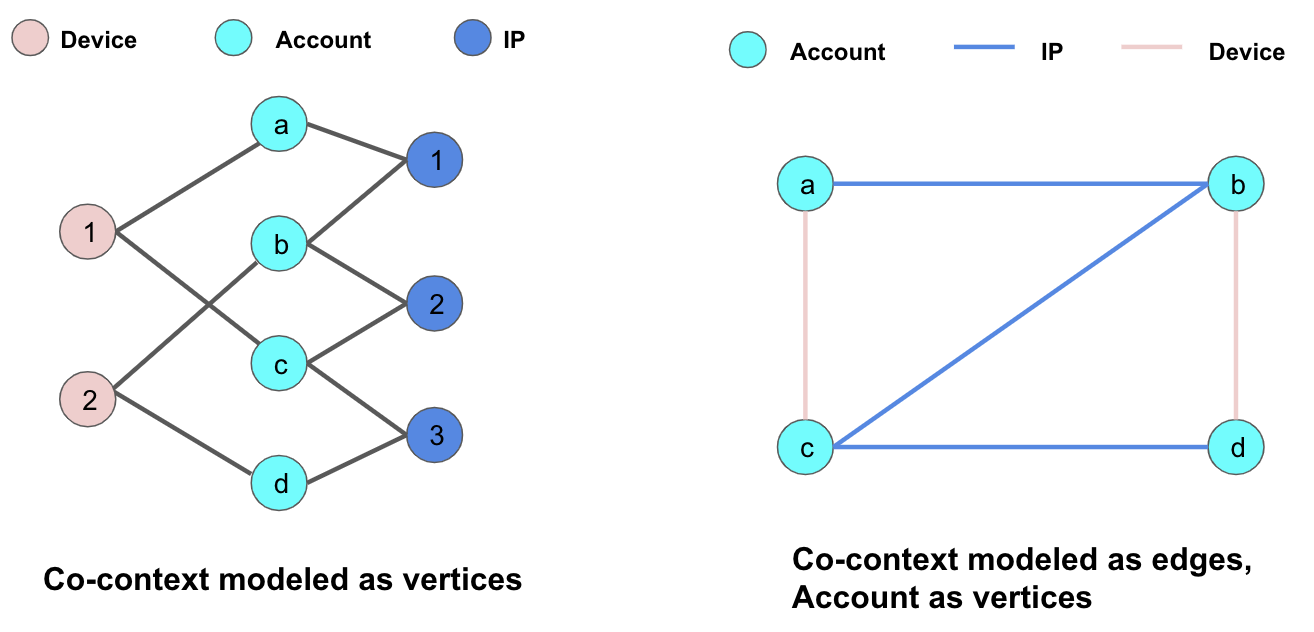}
  \caption{Co-Context Edge}
  \label{fig:concontext}
\end{figure}

\section{Application II}
In this application, we explore the implicit relationships between accounts and their shared resource.  We devise a time-sensitive co-context edge construction technique to construct the account graph. 

\subsection{Case Description}
One e-commerce website wants to setup a risk control system, which will be serving critical business steps to decide whether their current user's action post high risk to the business. The input to the risk control system would be the real-time risk control log, the entry of which is a risk control event. An event log usually contains event timestamp, type, event context (IP, location etc.) and an account. No explicit relationship between accounts exists in the event log. 

For each new risk control event, the system will digest the event information and update the risk scores for the affected accounts. Subsequently, any applications relying on the risk control system can real-time get the risk score for their current user, and take action when high risk score is returned.  

\subsection{IP Co-context Graph}
In this graph construction, the input data does not contain explicit relationships between accounts. To connect accounts, we use co-context edge construction technique to explore the share resource behaviors of the fraudulent accounts. 

We first describe the co-context edge concept and then show how we apply it to our current data set.

\smallskip
\noindent \textbf{Co-context Edge.}
How do we connect accounts from an account activity log when there is no explicit relationship? The answer is to use co-context. A co-context abstracts a shared resource as a context, and accounts sharing the same context will be connected. 

For example, an account event log contains the device id used, the account and the IP information. One way to construct a graph from the log entry is shown on the left of Figure \ref{fig:concontext}, where we create a vertex for each account, device and IP respectively, and connect an account and an IP, and connect an account and a Device when they appear in the same log entry. There are certain drawbacks of this approach. (1) It does not handle dynamic IP assignment issue, since an IP can be mixed used by both good and bad accounts. (2) It needs two hops from one account to reach the other account: one hop is from the account to IP, the other hop is from the IP to another account. This added hop incurs computational overhead. (3) It is hard to define the strength between two accounts over the two-hop model. 

We can use co-context concept to connect accounts. In this modeling approach, we collapse the above two-hop into one context edge.  For shared resource entities that co-occur with a pair of accounts, we create edges to represent each shared resource. On the right graph of Figure \ref{fig:concontext}, when two accounts share the same IP, we create an edge to connect them. The blue edges between a and b, b and c, and c and d indicate each pair shares the same IP. The pink edges between a and c, and b and d indicate that each pair shares the same device. We call this modeling method co-context edge modeling method. It has the following advantages:
\begin{itemize}
    \item Connecting account vertices directly by the contextual edge, it is easier to define the link strength between the end points.
    \item It is easier to merge edges between accounts. We can summarize multiple co-context edges into one since they share the same end points.
    \item Another advantage is that we just need 1-hop to traverse from one account to another, achieving the best performance. 
\end{itemize}

\begin{figure}[ht]
  \centering
  \includegraphics[width=1 \linewidth]{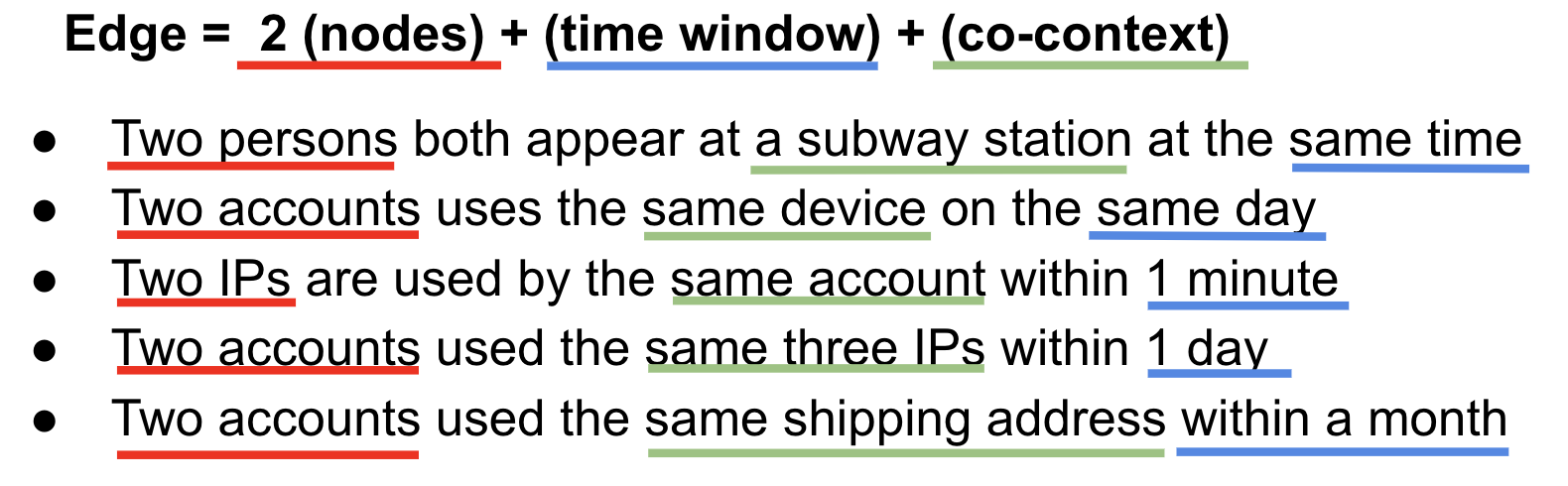}
  \caption{More Co-Context Examples.}
  \label{fig:cocontex-ex}
\end{figure}

\smallskip
\noindent \textbf{Time Constraint and Timestamp of Co-context Edge.} Fraudsters manipulate a common set of resources (IPs, Devices etc.). Their generated events show vicinity along the time dimension when they re-use the shared resources. Seeing this, we can add a time window constraint to further reduce the generated co-context edges. Imagine the following, an IP address was used at 9:00am by user u1, and used again by user u2, and u3 at 15:00pm and 15:01pm respectively. Then, the IP edge strength between u1 and u2 is weaker than the IP edge strength between u2 and u3. 

To construct time-sensitive co-context edge, we can add a time window while sweeping the event log. For example, we can set the window size to be 30 seconds. In the aforementioned example, (u2, u3)'s time stamp difference is less than 30 seconds, they will form a co-context edge; we use the latter timestamp of the two endpoints as the edge's creation timestamp. In the (u2, u3)'s case, the created edge has timestamp 15:01pm. On the other hand, (u1, u2)'s will not form a co-context edge as they do not fall within a time window.

Figure \ref{fig:cocontex-ex} shows more examples that we can use this modeling approach to create edges.

\begin{figure}[ht]
  \centering
  \includegraphics[width=1 \linewidth]{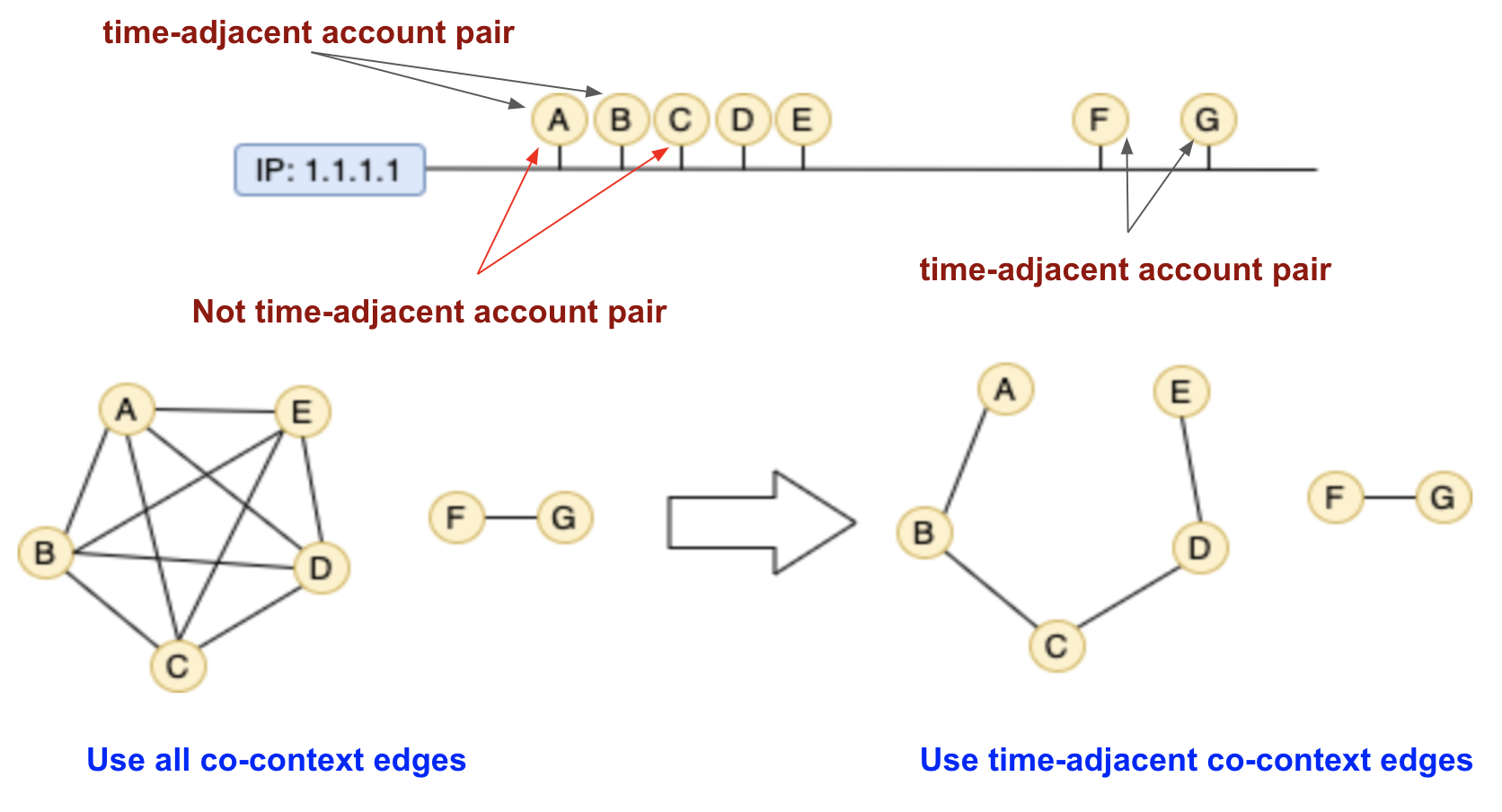}
  \caption{Reduce Co-context Edge Count}
  \label{fig:simple}
\end{figure}

\smallskip
\noindent \textbf{Applying Co-context Edge.}
In this application, each log entry has the account, the IP it uses, and the event type and timestamp. We create the account graph using the co-IP context. The graph has an Account vertex type, and a un-directed shared\_ip edge type, connecting two Account vertices when they both use the same IP within a 30-second time window. 

When constructing the shared\_ip edges, instead of creating an edge per pair of accounts in the same time window, we only create an edge for time-adjacent account pairs. This implementation greatly reduces the total edge count, while still preserves the important distance information between any two accounts from a CC. It is  illustrated in Fig \ref{fig:simple}. 

\subsection{Analytical Queries}
The analytic query we use is to cluster accounts into CCs using the shared\_ip co-context edges.  After we find the account CCs, we profile each account CC using the CC size statistic. We record in each account as its properties the cc size and the update timestamp. These two materialized properties are used to answer real-time risk control request.

Using the CC size as the profiling statistics deserves further discussion. When fraudsters' programs manipulate accounts using their limited IPs, to maximize their profits of the targeted campaigns, they tends to maximize the amount of their account activities with the shared IP. It's very unlikely two distinct normal accounts of a website will share the same IP as the IPs are usually assigned according to the same geolocation or ISP. Even when two normal accounts do share an IP, the chances that they both visit the website at a small time window is very slim. As a result, with a small time window as the shared\_ip creation criteria, we can effectively capture the fraud account clusters.

There do exist scenarios where this method may not work. For example, for a national holiday such as Thanksgiving or Christmas, peoples living in the same neighborhood are more likely to visit the same website for online shopping on the same day. In such scenario, two normal accounts may share an IP within a small time window.

On the other hand, suppose the fraudsters buy some commercial proxy services to obtain a new IP per their account use to bypass this method, it will incur non-trivial cost for the fraudsters. Increasing the cost of fraudsters attack is another way to deter their efforts. 
\subsection{Experiment}

\noindent \textbf{Goal.} Our goal is to answer the following questions before deploying the risk control system to production.  

\begin{enumerate}
\item Is this method scalable in terms of storage and performance with respect to the size of the risk control log?
\item Can this system provide high QPS to serve as an operational risk control service 24x7?
\item How are the precision and recall affected by the con-context time window size?
\item What are the precision and recall? With what parameters setting?

\end{enumerate}

\noindent \textbf{Setup.} 
In this experiment, we use the e-commerce website’s 10-day risk control log from 2019-12-01 to 2019-12-09. The log contains 4344376 raw account events, and about 1.2 million accounts. The Tencent's T-Sec API  flagged 7.6k accounts as high risk accounts, which serves as our canonical truth when computing precison and recall. 

For all the experiments, we use a server with Intel(R) Core(TM) i7-7800X CPU@3.50GHz, 12 Cores and 64G RAM. 

To test the scalability of this method, we used the full 10 days data, and varied the window size of constructing co-context edges from 10s to 3600s. We varied the window size since with bigger co-context edge construction windows size, for the same input data, we will have bigger account graph, which helps us to test our system scalability with regard of account graph size. ßWe record the vertex and edge counts with respect to the window size. On the account graphs of different window size, we record the elapsed time to do one pass of CC update. 

To test the QPS, we ran two experiments over the full 10 days data, and fixed the window size to 3600s to stress test the worst situation. The first experiment tests the throughput of CC size lookup query by repeatedly sending concurrent REST requests to the graph database with no CC detection query running in the background. The second experiment tests the same lookup query QPS, but with the CC detection and account CC size update query in the background 24x7. We reported the QPS of each. 

To measure the time window size's influence on the algorithm accuracy, we compute the precision, recall and F score for each window size over the same 10-day log. 

To compute the precision and recall, we set the time window size to 30s, and the CC size threshold to 10. We show the charts with different values of the two parameters to explain why we pick them. The constructed account graph has 95,827 accounts, connected by 92,413 shared\_ip edges, excluding those accounts that has no edges. We run the un-directed  CC detection algorithm and update the CC size of each account. 

\noindent \textbf{Result.} 

\begin{figure}[ht]
  \centering
  \includegraphics[width=1.1 \linewidth]{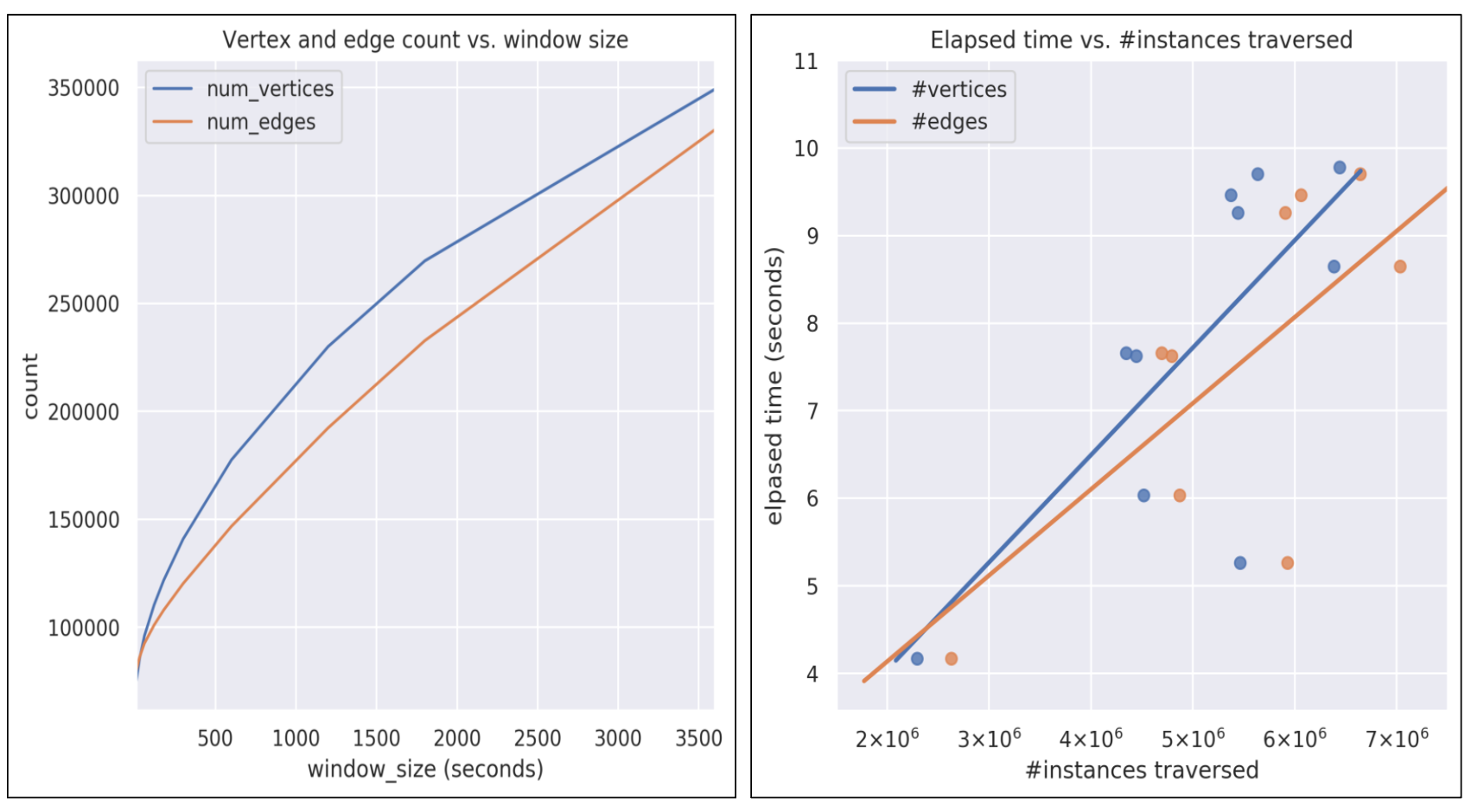}
  \caption{Scalability Test.}
  \label{fig:scal}
\end{figure}

In Fig. \ref{fig:scal}, the left chart shows that as the window size increases, the total account number in the account graph is going up sub-linearly, while the total number of co-context edge is going up linearly. Usually, as the vertex number goes up, the edge number exponentially goes up. In our implementation, since we only constuct co-context edge with time-adjacent accounts, we achieve linear scalability of the edge count, which resulted linear storage scalability of the size of the constructed account graph.  In Fig. \ref{fig:scal}, the right chart shows the CC computation time v.s. each account graph's (created by different window size) vertex and edge count. The elapsed time to compute CC for the entir graph is linear with respect to the graph element size, showing good performance scalability of this method. 

\begin{table}[]
\begin{tabular}{|l|l|l|}
\hline
\textbf{Measurement} & \textbf{No bkgrd CC alg} & \textbf{CC bkgrd alg} \\ \hline
\textbf{95\% pct}    & 25ms                         & 31ms                      \\ \hline
\textbf{QPS}         & 14.57K                       & 11.89k                    \\ \hline
\end{tabular}
\caption{QPS testing.}
\label{Tab:qps}
\end{table}

In Table \ref{Tab:qps}, we show the 95 percentile of  the QPS benchmark with or without the CC algorithm running in the background. Without the CC algorithm running in the background, 95\% of the CC size lookup  query can finish within 25ms. With the CC algorithm running, we achieve 31ms upper limit for the 95\% of query. This shows excellent  real-time response performance, which is critical for risk control system, since the sooner it discovers fraudulent account, the easier for the client to save the economic loss of the attack. The two QPS 14.57k v.s. 11.89k are both above most of the risk control system requirement. For our client, they expect 5k will be sufficient for their business operation.

\begin{figure}[ht]
  \centering
  \includegraphics[width=1.1 \linewidth]{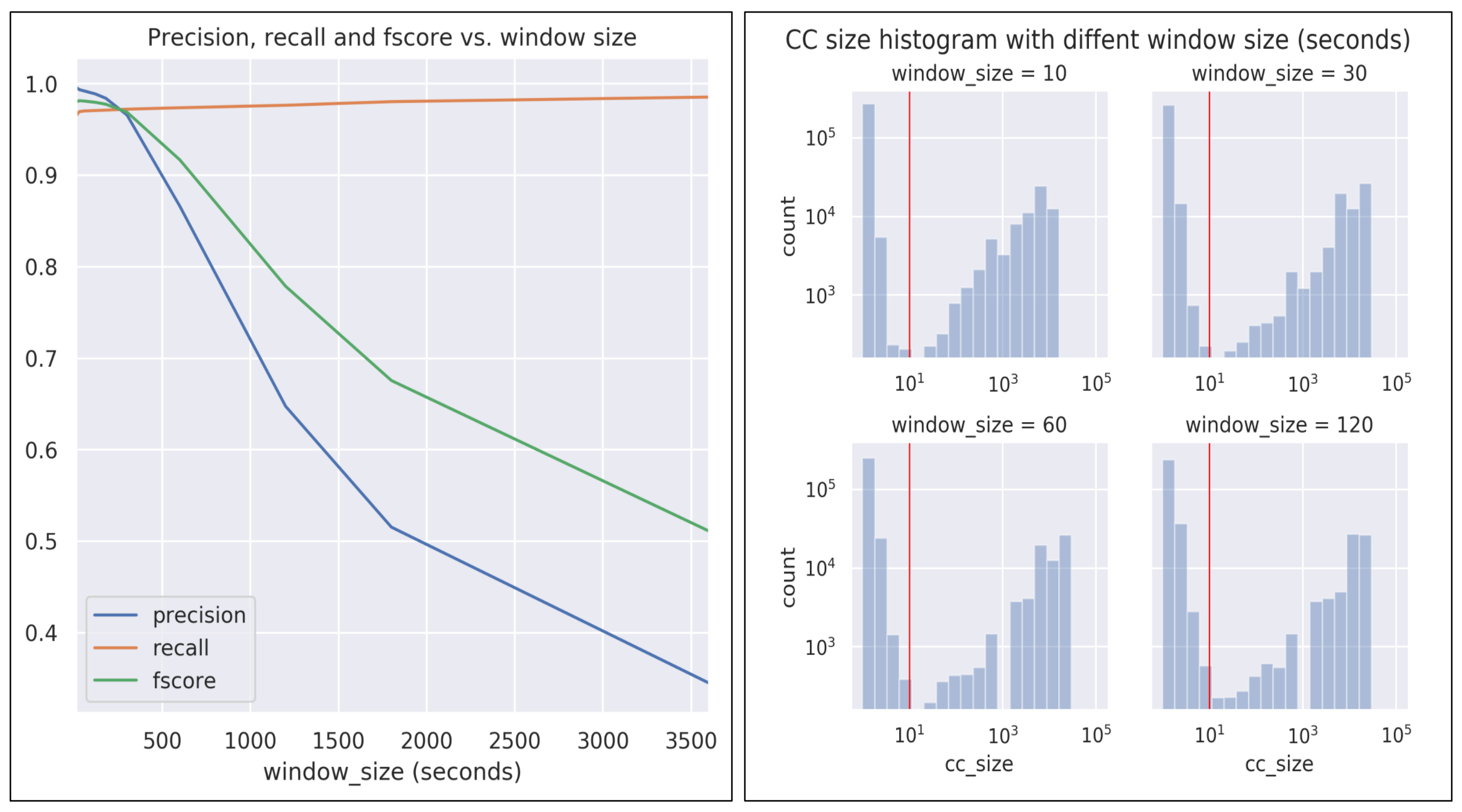}
  \caption{Picking Parameters. }
  \label{fig:param}
\end{figure}

Finally, we discuss the relationship between parameters and precision and recall. In Fig. \ref{fig:param}, the left chart shows with larger window size to construct the shared\_ip edge, the system's precision drops significantly. Good precision and recall are achieved when the window size is less than 100s. The reason the false positive increases is that with a larger window, the likelihood of two normal account share the same IP is increased, thus causing many false alarms. In Fig. \ref{fig:param}, the right chart shows the CC size histograms of different window sizes. The x-axis is the CC size, and the y-axis is the CC count.  The red line marks the CC size of 10. For this data set, we can pick any CC size threshold from 10 to 100 without changing the fraud detection rate. We pick 10 as our final production parameter as it provides a good trade-off between precision and recall, and the sensitivity of our risk control system. The real deployment can fine tune this parameter to adapt to the production usage pattern.     

\begin{enumerate}
\item Precision and Recall:
\item Scalability:
\item Choosing the CC size threshold:
\item QPS. 
\end{enumerate}

 The framework reports 70,086 accounts  as risky accounts. Among them, 69532 also flagged as risky by the cloud service A. Note that the service A flagged total 7.6k accounts as risky accounts. The 6k accounts missed by the co-context graph is by design, as they are not part of the constructed graph. Even without having the opportunity to score the missed 6k accounts, the account graph has a detection rate of 91.4\% (69532/7.6k), which is pretty good. 

\subsection{Production Deployment.}

After the succesful PoC, our client has deployed the delivered system to production since April 2020. In the deployment, when construction shared\_ip edges, we used the window size of 3600s. On each edge, we store its creation time and the time delta of the two account events it connects. When computing CC, our GSQL query filters those edges that are created beyond the latest 7 days. Also, by filtering edges based on its time delta property, we can apply any window size from 1s to 3600s to our production system. 
Since GSQL is installed as a stored procedure and invoke by REST end point, we can dynamically tune our algorithms with the window size to adapt to the real log stream changes. 

\section{Related Work}
PathSim \cite{metapath} explores the MetaPath on existing static heterogeneous network to compute the similarity of two nodes. Our co-context edge is real-time constructed for a dedicate risk control account graph. HitFraud \cite{hitfraud} conducted semi-supervised learning by extracting graph features from the existing transaction graph connected by MetaPath\cite{metapath}, and these features are used with other labeled data to train a transaction classifier. Our approach is unsupervised and builds a dedicate dynamic account graph from the log, and only add to the graph the shared resources vertices. We study the machine behavior of account CCs, not individual account. HGsuspector \cite{hgsuspector} decomposes directed offline property graph into a set of bi-party graphs, account - x (ip, device, etc.) , then it scores each connected bi-party graph, and report anomaly if a threshold is met. The drawback is that it is not explainable and it passively takes an existing graph. Our approach regards  constructing a dedicate account graph in real-time as the first step is important, and only adding risk-control relevant edges between accounts.  The framework explores the program mechanic pattern and shared resource pattern on the account CCs, and the results are explainable.  

\section{Conclusion}
We have presented two risk control systems to detect collective fraud. Both system are implemented by GSQL and powered by TigerGraph database. We adopt the approach to build a dedicate account graph to detect fraud. This is in contrast to detect fraud on existing heterogeneous network data, where many intermediate links and vertices existed between accounts. The account graph explores the machine behavior and the finite shared resource characteristic by profiling account CCs. Both applications demonstrate the practical advantage of building a risk control system using a graph database. 

\bibliographystyle{ACM-Reference-Format}
\bibliography{sample-base}


\end{document}